\documentclass[journal=jacsat,manuscript=communication,twoside,layout=two-column]{achemso}
\usepackage{amsmath,amssymb,units,txfonts}
\usepackage{amsfonts,times}
\usepackage{graphicx,multirow}
\usepackage{helvet,wasysym,bm}
\usepackage[bookmarks=true,colorlinks=true,urlcolor=blue,linkcolor=blue,citecolor=blue]{hyperref} 
\usepackage[usenames,dvipsnames]{color}
\usepackage{colortbl,cuted}

\definecolor{JPCCBlue}{RGB}{34,80,169}
\definecolor{JACSBlue}{RGB}{40,50,128}
\definecolor{abstractcolor}{RGB}{255,243,201}
\makeatletter\newenvironment{abstractbox}{
   \begin{lrbox}{\@tempboxa}\begin{minipage}{0.975\columnwidth}}{\end{minipage}\end{lrbox}
   \colorbox{abstractcolor}{\usebox{\@tempboxa}}
}\makeatother

\renewcommand*\authorfont{\flushleft}
\renewcommand*\affilfont{\mdseries\flushleft}
\renewcommand*\titlesize{\Large}
\renewcommand*\titlefont{\flushleft\sf\bfseries}
\def\refname{\sffamily\color{JACSBlue}\Large$\blacksquare$\normalsize\ REFERENCES}
\usepackage{titlesec}
\titleformat{\section}{\bfseries\sffamily\color{JACSBlue}}{\thesection.~}{0pt}{}
\titleformat{\subsection}[runin]{\bfseries\sffamily\normalsize}{\indent\thesubsection.~}{0pt}{}[.]
\titlespacing{\subsection}{0pt}{0pt}{*1}
\titleformat{\subsubsection}{\bfseries\sffamily\normalsize}{\thethesubsection.~}{0pt}{}
\titlespacing{\subsubsection}{0pt}{0pt}{*0}

\title{Level alignment of a prototypical photocatalytic system:\\Methanol on TiO$_{\textbf{{2}}}$(110)}

\author{Annapaoala Migani}
\email{amigani@cin2.es}
\affiliation[UPV/EHU]{\footnotemark[2]{\ } Nano-Bio Spectroscopy Group and ETSF Scientific Development Center, Departamento de F{\'{\i}}sica de Materiales, Centro de F{\'{\i}}sica de Materiales CSIC-UPV/EHU-MPC and DIPC, Universidad del Pa{\'{\i}}s Vasco UPV/EHU, E-20018 San Sebasti\'{a}n, Spain}
\alsoaffiliation[CIN2]{\newline\footnotemark[3]{\ } CSIC - Consejo Superior de Investigaciones Cientificas, ICN2 Building, E-08193 Bellaterra (Barcelona), Spain}
\author{Duncan J. Mowbray}
\email{duncan.mowbray@gmail.com}
\author{Amilcare Iacomino}
\affiliation[UPV/EHU]{\footnotemark[2]{\ } Nano-Bio Spectroscopy Group and ETSF Scientific Development Center, Departamento de F{\'{\i}}sica de Materiales, Centro de F{\'{\i}}sica de Materiales CSIC-UPV/EHU-MPC and DIPC, Universidad del Pa{\'{\i}}s Vasco UPV/EHU, E-20018 San Sebasti\'{a}n, Spain}
\author{Jin Zhao}
\affiliation[USTC]{\newline\footnotemark[5]{\ } Physics Department, Hefei National Laboratory for Physical Sciences at Microscale, University of Science and Technology of China, Hefei, Anhui 230026, China}
\author{Hrvoje Petek}
\affiliation[UP]{\newline\footnotemark[4]{\ } Department of Physics and Astronomy, University of Pittsburgh, Pittsburgh, Pennsylvania 15260, USA}
\author{Angel Rubio}
\email{arubio@ehu.es}
\affiliation[UPV/EHU]{\footnotemark[2]{\ } Nano-Bio Spectroscopy Group and ETSF Scientific Development Center, Departamento de F{\'{\i}}sica de Materiales, Centro de F{\'{\i}}sica de Materiales CSIC-UPV/EHU-MPC and DIPC, Universidad del Pa{\'{\i}}s Vasco UPV/EHU, E-20018 San Sebasti\'{a}n, Spain}

\begin{document}
\maketitle
\noindent{\color{JACSBlue}{\rule{\columnwidth}{0.5pt}}}
\begin{abstractbox}
\noindent\textbf{\color{JACSBlue}{ABSTRACT:}}
Photocatalytic and photovoltaic activity depends on the optimal alignment of electronic levels at the molecule/semiconductor interface.   Establishing level alignment experimentally is complicated by the uncertain chemical identity of the surface species.
We address the assignment of the occupied and empty electronic levels for the prototypical photocatalytic system of methanol on a rutile TiO$_{\text{2}}$(110) surface.
Using many-body quasiparticle (QP) techniques we show that the frontier levels measured in ultraviolet photoelectron and two photon photoemission spectroscopy experiments can be assigned with confidence to the molecularly chemisorbed methanol, rather than its decomposition product, the methoxy species.  
We find the highest occupied molecular orbital (HOMO) of the methoxy species is much closer to the valence band maximum, suggesting why it is more photocatalytically active than the methanol molecule. 
We develop a general semi-quantitative model for predicting many-body QP energies 
based on the appropriate description of electronic screening within the bulk, molecular or vacuum regions of the wavefunctions at molecule/semiconductor interfaces.
\end{abstractbox}
\noindent{\color{JACSBlue}{\rule{1.\columnwidth}{0.5pt}}}

\maketitle

\def\bigfirstletter#1#2{{\noindent
    \setbox0\hbox{{\color{JACSBlue}{\Huge{#1}}}}\setbox1\hbox{#2}\setbox2\hbox{(}%
    \count0=\ht0\advance\count0 by\dp0\count1\baselineskip
    \advance\count0 by-\ht1\advance\count0 by\ht2
    \dimen1=.5ex\advance\count0 by\dimen1\divide\count0 by\count1
    \advance\count0 by1\dimen0\wd0
    \advance\dimen0 by.25em\dimen1=\ht0\advance\dimen1 by-\ht1
    \global\hangindent\dimen0\global\hangafter-\count0
    \hskip-\dimen0\setbox0\hbox to\dimen0{\raise-\dimen1\box0\hss}%
    \dp0=0in\ht0=0in\box0}#2}
\bigfirstletter
Molecular energy levels are strongly renormalized when molecules are brought into contact with surfaces \cite{RenormalizationLouie,*JuanmaRenormalization1}.  The energy positions of the frontier highest occupied and lowest unoccupied molecular orbitals (HOMO and LUMO) of the adsorbate with respect to the valence band maximum and conduction band minimum (VBM and CBM) of photocatalytic substrates define the potentials for electron transfer across a molecule/semiconductor interface.  
Photoelectron spectroscopy can accurately determine the alignment of the frontier orbitals of the adsorbate with respect to the substrate bands provided that the chemical state of the chemisorbed molecule is known. The chemical assignment and correct description of the molecule-photocatalyst interaction require theory to accurately predict the electronic structure of the coupled system.

We consider the electronic structure of methanol chemisorbed intact or in its partially dissociated methoxy form on the stoichiometric rutile TiO$_{\text{2}}$(110) surface. Methanol is well established as a sacrificial hole scavenger in the photocatalytic splitting of H$_{\text{2}}$O by UV light excitation of TiO$_{\text{2}}$ nanocolloids \cite{Kawai1980,FujishimaReview}. Experimentally, the electronic structure and photocatalytic activity of methanol on the single crystal rutile TiO$_{\text{2}}$(110) surface under ultra-high vacuum (UHV) conditions has been investigated by ultraviolet, X-ray, and two photon photoelectron spectroscopy (UPS, XPS, and 2PP) \cite{Onishi198833,Weixin,PetekScienceMethanol,MethanolSplitting2010}, scanning tunnelling microscopy \cite{MethanolSplitting2010,HendersonReview,Henderson2012} (STM), and mass spectrometric analysis of reaction products \cite{Henderson2011,*Friend2012,MethanolPhotocatalysis}. These experiments have reached contradictory conclusions regarding whether the empty ``wet electron'' level \cite{Onda20052005} that is observed in 2PP spectra of methanol covered TiO$_{\text{2}}$ surfaces should be assigned to the methanol or methoxy species \cite{PetekScienceMethanol, MethanolSplitting2010}. Furthermore, although it is clear that the methoxy species is more photocatalytically active than the methanol molecule, there is still debate as to whether the decomposition of methanol to formaldehyde and methyl formate is initiated by the thermal \cite{Henderson2012,Henderson2011,*Friend2012} or photocatalytic \cite{Weixin,MethanolSplitting2010,MethanolPhotocatalysis} decomposition of methanol.  Establishing the mechanism of this prototypical photocatalytic process
critically depends on the correct attribution of photoemission spectra and determination
of the energy alignment of the HOMO and LUMO of the methanol and methoxy species.

\renewcommand{\thefootnote}{}
\footnotetext{\hspace{-0.5cm}This document is the unedited Author's version of a Submitted Work that was subsequently accepted for publication in Journal of the American Chemical Society, copyright \copyright American Chemical Society after peer review. To access the final edited and published work see \href{http://dx.doi.org/10.1021/ja4036994}{http://dx.doi.org/10.1021/ja4036994}.}

Previous studies have shown electronic structure calculations based on a Hartree-Fock approach adequately describe the molecular energy levels in vacuum \cite{RenormalizationLouie,*JuanmaRenormalization1}. For photocatalytic systems, covalent bonds between the molecular overlayer and the substrate inevitably involve significant hybridization of the interacting levels, which is well described at the level of density functional theory (DFT).  As the molecule approaches the substrate, the mutual polarization of their charge distributions, i.e.~screening, also renormalizes the energy of the molecular levels. A proper treatment of the inhomogeneous screening by the environment requires the use of many-body quasiparticle (QP) techniques \cite{AngelGWReview, *Galli, *GiustinoPRL2012}.    The inhomogeneous screening must be taken into account to obtain even a qualitative understanding of the energy level alignment. \cite{AngelGWReview, *Galli, *GiustinoPRL2012}.

We carry out many-body QP calculations at the $G_0W_0$ 
level and introduce the $scGW1$ self consistent approach described in the 
Supporting Information \cite{KresseG0W0,*KresseGW,*KressescGW,*CarusoPRB2012}. These are based on DFT calculations using a generalized gradient approximation to the exchange and correlation (xc)-functional (PBE) \cite{PBE}.  We have also performed self consistent QP calculations at the $scGW$ and $scGW_0$ levels, and DFT calculations using a hybrid xc-functional (HSE)
\cite{HSE}.  PBE, HSE, $scGW$, and $scGW_0$ calculations all fail to describe  even qualitatively the level alignment of methanol on TiO$_{\text{2}}$(110) (see Supporting Information). However, from $G_0W_0$ and $scGW1$ calculations we obtain the correct level alignment for both the occupied and empty molecular levels at the interface.  This enables us to conclude that the molecular structures measured in UPS \cite{Onishi198833} and 2PP \cite{Onda200532} experiments can mostly be attributed to intact methanol molecules on TiO$_{\text{2}}$(110).   For the partially dissociated methanol layer we find the HOMO of the methoxy species is nearer the VBM. This more favourable HOMO alignment may explain  why the methoxy species is more photocatalytically active than the methanol molecule.

Of more general significance, we find that the many-body QP corrections to the DFT levels' energies, i.e.\ $G_0W_0$ QP energy shifts, are correlated quantitatively with the fraction of the level's density within the bulk, molecular, and vacuum regions.  In fact, the differences in these corrections directly reflect the inhomogeneous electronic screening at the interface. Therefore, the bulk, molecular or vacuum character of the wavefunctions acts as a ``descriptor'' for the QP energy shifts.  Based on this finding, we are able to construct a general model that quantitatively describes the QP corrections to the electronic energy levels at a molecule/semiconductor interface.

The adsorption of methanol on rutile TiO$_{\text{2}}$(110) is highly inhomogeneous due to the similarity in energy of several configurations.  Methanol can chemisorb intact through its O atom to a coordinately unsaturated (cus) Ti site  of the substrate, and by forming a hydrogen bond between the H atom of its OH group and the surface bridging O atom. It can also dissociate by transferring the H atom to the surface through the hydrogen bond to leave a methoxy species at the Ti cus site. The degree of dissociation, and whether it occurs thermally or photocatalytically is unknown.  On the one hand, the primary photocatalytically active species may be the methanol molecules, which upon excitation of TiO$_{\text{2}}$ with UV light deprotonates to form a methoxy species, as identified by single molecule-resolved STM \cite{MethanolSplitting2010} and XPS \cite{Weixin}.  On the other hand, it may be the methoxy species, which is 
 present due to thermal decomposition of a methanol molecule, and is photo-oxidized to formaldehyde and methyl formate, as suggested in the recent studies of TiO$_{\text{2}}$ photocatalysis \cite{Henderson2012,Henderson2011,*Friend2012}.

\begin{figure}[!t]

\includegraphics[width=0.9\columnwidth]{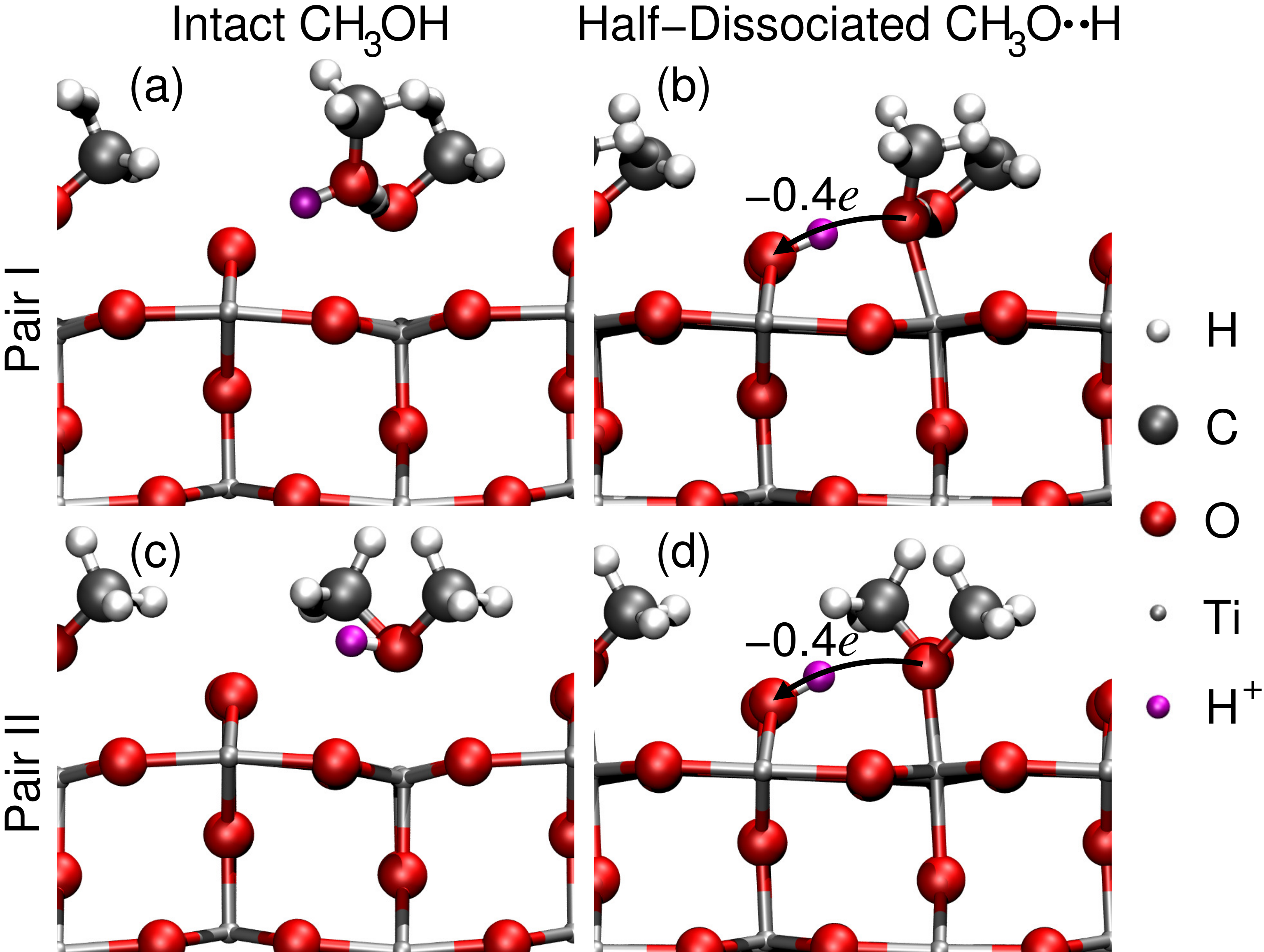}
\caption{The atomic structure of methanol monolayers on TiO$_{\text{2}}$(110). Geometries for (a,c) intact and (b,d) half-dissociated pair I (a,b) and pair II (c,d) methanol monolayers.  The transferred proton is marked in magenta, along with the accompanying charge transfer of 0.4 electrons.
}\label{fgr:0}
{\color{JACSBlue}{\rule{\columnwidth}{1.0pt}}}
\end{figure}

The relative stabilities and structures of the fully intact, 50\% dissociated, and 100\% dissociated methanol monolayers have been calculated in Ref.\ \citenum{Jin}\nocite{Jin}. We performed $G_0W_0$ QP calculations for the interfacial electronic structure of the four most stable methanol monolayers (Fig.~\ref{fgr:0}), which includes two ``intact'' methanol monolayers, and their ``half-dissociated'' counterparts. In these structures half the methanol molecules have an intermolecular hydrogen bond, while the other half have an interfacial hydrogen bond.  Through this interfacial hydrogen bond proton transfer and an accompanying charge transfer of 0.4 electrons occur from the oxygen of the methanol to the nearest surface bridging oxygen (Fig.~\ref{fgr:0}). 

Bridging oxygen vacancies, sub-stoichiometricity of the bulk, and interstitial Ti atoms are not accounted for in these models. Nevertheless, the simulated and measured XPS \cite{Weixin} C1\textit{s} (-0.66 vs.\ -0.6 eV) and O1\textit{s} (-1.94 vs.\ -1.7~eV) core-level shifts are in semi-quantitative agreement for the most stable intact methanol monolayer and its half-dissociated counterpart (pair I). 
This demonstrates that these structures are realistic models for UHV experiments.
Here, we focus on the ``initial state'' electronic structure of the interface prior to photon irradiation under UHV conditions as the HOMO level alignment may change upon creation of a hole or due to a solvent.\cite{Prezhdo2013JACS}

\begin{figure}[!t]
\includegraphics[width=\columnwidth]{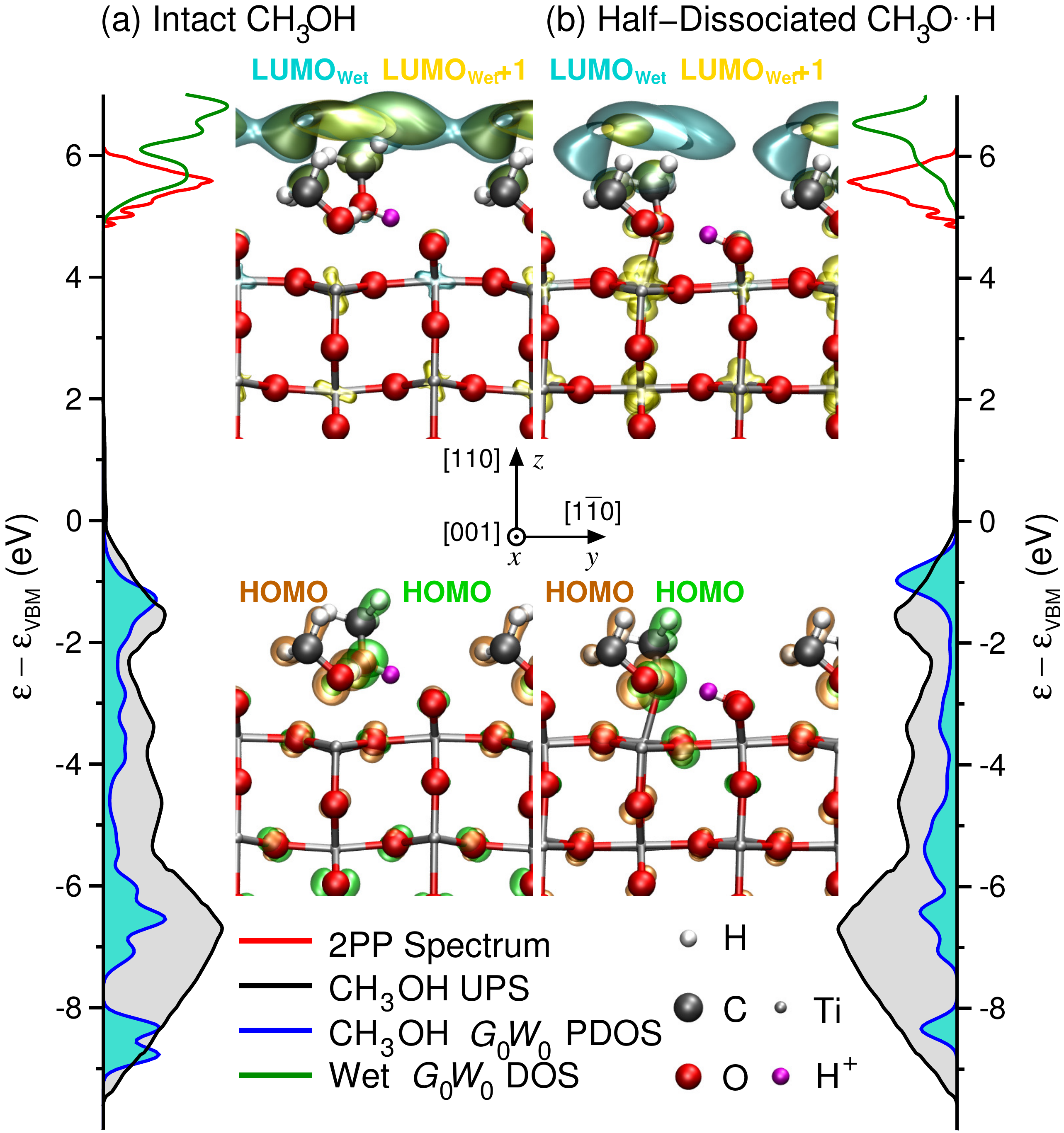}
\caption{
The electronic structure of methanol monolayers on TiO$_{\mathbf{2}}$(110). CH$_{\text{3}}$OH projected (blue), and Wet (green) DOS  computed with $G_0W_0$ for the depicted (a) intact  and (b) half-dissociated pair I structures, are compared with the UPS \cite{Onishi198833}  (black) and  2PP spectrum (red) \cite{Onda200532}. Filling denotes occupied levels. Energies are relative to the VBM, $\varepsilon_{\textrm{VBM}}$. The H atom undergoing proton transfer is marked in magenta. HOMO  for the (green) deprotonating methanol molecule, and (orange) methanol molecule that is hydrogen bonded to it, are shown together below with the LUMO$_{\textrm{Wet}}$ (cyan) and LUMO$_{\textrm{Wet}}$+1 (yellow) above.  }\label{fgr:1}
{\color{JACSBlue}{\rule{\columnwidth}{1.0pt}}}
\end{figure}

Fig.~\ref{fgr:1} shows the electronic structure for pair I.  First, we focus on the highest energy peak in the UPS at $\varepsilon_{\textit{peak}}^{\textrm{UPS}}\approx -1.55$~eV relative to the VBM ($\varepsilon_{\textrm{VBM}}$). By comparing the projected density of states (PDOS) onto the methanol layer with UPS measurements \cite{Onishi198833} of CH$_{\text{3}}$OH/TiO$_{\text{2}}$  in Fig.~\ref{fgr:1}(a), we find the corresponding peak in semi-quantitative agreement with the intact structure, which is $\sim 0.3$~eV closer to the VBM. This PDOS feature corresponds to  distinct HOMOs localized on each methanol molecule within the unit cell, as  depicted in green or orange in Fig.~\ref{fgr:1}. These are predominantly non-bonding O2$p$ orbitals, with some C--H$\sigma$ and Ti3$d$ character.  The weak hybridization of the HOMO orbitals with the substrate is reflected in the relative narrowness of the primary peaks. The PDOS for the half-dissociated structure shown in Fig.~\ref{fgr:1}(b) is shifted $\sim 0.6$~eV closer to the VBM than the UPS data.  The comparison with UPS experiments suggests that it is mostly the intact methanol layer which is measured at $\sim$298 K\cite{Onishi198833}.

Next, turning to the unoccupied molecular levels in Fig.~\ref{fgr:1}, we find that they  have primarily two dimensional (2D) $\sigma*$ character associated with the methanol C--H bond, with weight above H atoms outside the molecular layer \cite{PetekScienceMethanol}. These are the ``wet electron'' levels \cite{Onda20052005}, which give an intense experimental peak at $\varepsilon_{\textit{peak}}^{\textrm{2PP}}\approx 5.58$~eV in 2PP spectra \cite{Onda200532}.  The wet electron density of states (Wet DOS)  at the $G_0W_0$ level  is also in better agreement with experiment for the intact structure rather than the half-dissociated one, as is evident in Fig.~\ref{fgr:1}.  The LUMO$_{\textrm{Wet}}$ and LUMO$_{\textrm{Wet}}+1$ levels, which are located at the onset of the Wet DOS spectrum, are shown in Fig.~\ref{fgr:1}.

\begin{table}[!t]
\caption{\rm\textbf{Energies of the highest/lowest peaks $\bm{\varepsilon_{\textit{peak}}^{\textrm{PDOS/Wet}}}$ in eV relative to the VBM $\bm{\varepsilon_{\textrm{VBM}}}$, for the CH$_{\text{3}}$OH PDOS and Wet DOS, and differences $\bm{\Delta \varepsilon}$ from the UPS and 2PP measurements, respectively, for the intact (CH$_{\text{3}}$OH) and half-dissociated (CH$_{\text{3}}$O$\bm{\cdot\cdot}$H) pair I and II structures.}
}\label{tbl:1}
\footnotesize
\begin{tabular}{ccccccc}
\multicolumn{6}{>{\columncolor[gray]{0.9}}c}{ }\\[-3mm]
\rowcolor[gray]{0.9}
 Level of &  Molecular &  Pair & $\varepsilon_{\textit{peak}}^{\textrm{PDOS}}$ &  $\Delta \varepsilon^{\textrm{UPS}}$&  $\varepsilon_{\textit{peak}}^{Wet}$ &  $\Delta \varepsilon^{\textrm{2PP}}$\\
\rowcolor[gray]{0.9}
 Theory 	 &  Layer & &  (eV) &  (eV) &  (eV) &  (eV)\\
\multirow{4}{*}{$G_0W_0$}
& 
\multirow{2}{*}{CH$_{\text{3}}$OH}
 	&I & -1.29 & +0.26 & +5.68 & +0.10\\
& 	&II&-1.44&+0.11&+5.19&-0.39\\
& 
\multirow{2}{*}{CH$_{\text{3}}$O$\cdot\cdot$H}
 	&I & -0.97 & +0.57 & +5.98 & +0.40\\
& 	&II&-0.95&+0.60&+6.13&+0.55\\
$scGW1$ 	& 
\multirow{1}{*}{CH$_{\text{3}}$OH} 
		 &I& -1.70 & -0.15 & +5.62 & +0.04\\
UPS/2PP	&& & -1.55$^{\textit{a}}$
& ---   & +5.58$^{\textit{b}}$
& --- \\
\multicolumn{2}{l}{$^{\textit{a}}$Ref.~\citenum{Onishi198833}\nocite{Onishi198833}, $^{\textit{b}}$Ref.~\citenum{Onda200532}\nocite{Onda200532}}
\end{tabular}
{\color{JACSBlue}{\rule{\columnwidth}{1.0pt}}}
\end{table}

Our $G_0W_0$ results favour the assignment of both the UPS and 2PP measurements  to the intact methanol overlayers ($\Delta \varepsilon^{\textrm{UPS}} \approx +0.26$~eV, $\Delta \varepsilon^{\textrm{2PP}} \approx +0.10$~eV in Table \ref{tbl:1}).  This is fully supported by the PDOS and Wet DOS for the intact and half-dissociated structures of the second most stable pair II \cite{Jin}.  Compared to pair I, the molecule-surface H bond is weakened, and the two methyl groups are reoriented away from each other for pair II (Fig.~\ref{fgr:0}). As shown in Table~\ref{tbl:1}, the two intact structures have similar PDOS and Wet DOS peak energies.  The correspondence is even closer for the two half-dissociated structures. Overall, the shape of the spectra are quite similar in both cases. This means the spectral assignment to intact rather than half-dissociated structures is robust against such differences in orientation within the molecular overlayer.

$G_0W_0$ calculations only provide the QP eigenenergies, as the QP wavefunctions and vacuum level are not computed. Therefore,
the influence of the screening on the vacuum level and wavefunctions is not directly available at the $G_0W_0$ level.  
To quantify this effect we must go beyond $G_0W_0$ to the self consistent $GW$ level.  

\begin{figure}
{\color{JACSBlue}{\rule{\columnwidth}{1.0pt}}}
\includegraphics[height=1.28in]{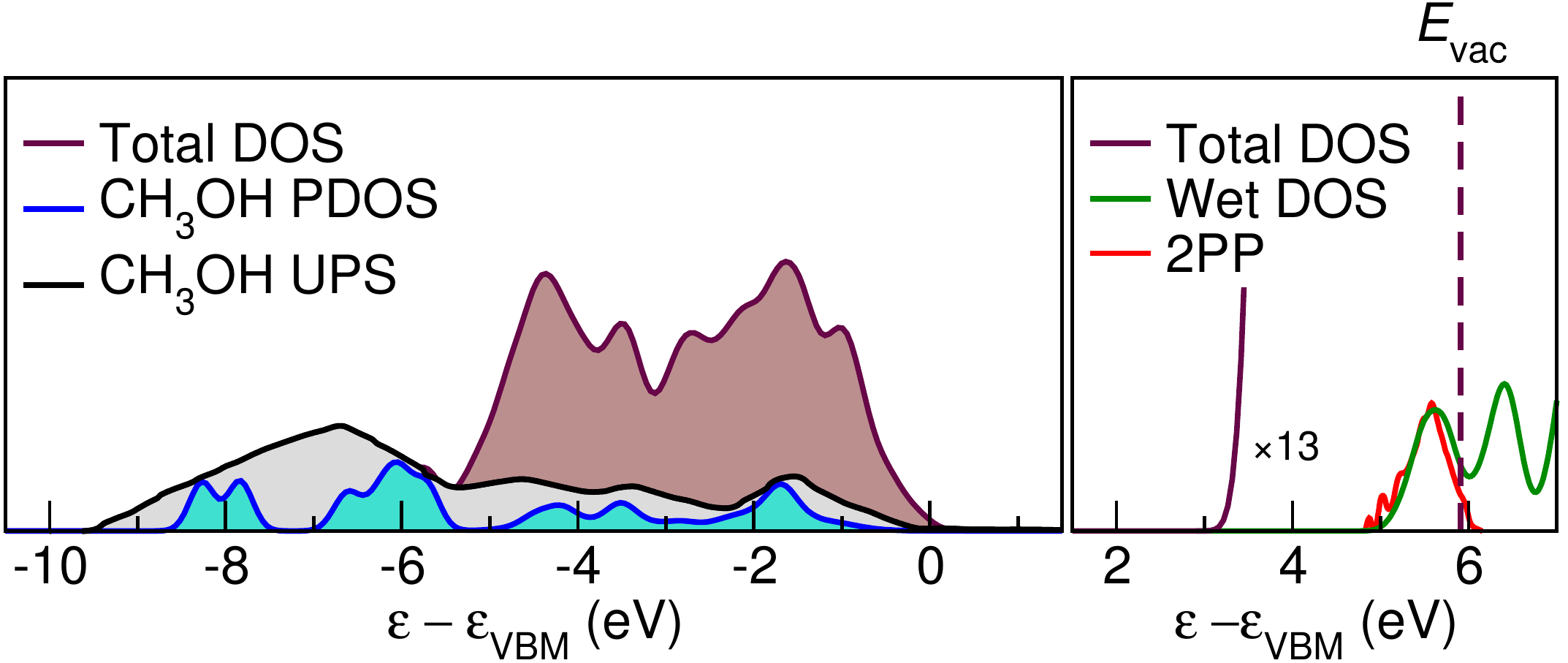}
\caption{
Total (maroon), CH$_{\text{3}}$OH projected (blue), and Wet (green) DOS  computed with $scGW1$ for the intact structure of pair I. The calculated CH$_{\text{3}}$OH PDOS and Wet DOS are compared with the UPS \cite{Onishi198833}  (black) and 2PP spectrum \cite{Onda200532} (red). Filling denotes occupied levels. Energies are relative to the VBM, $\varepsilon_{\textrm{VBM}}$. The dashed vertical line indicates the vacuum level $E_{\textit{vac}}$.}\label{fgr:3}
{\color{JACSBlue}{\rule{\columnwidth}{1.0pt}}}
\end{figure}

To maintain the accurate $G_0W_0$ description of the spectra while also describing the vacuum level and QP wavefunctions via the self consistent $GW$ procedure, we introduce the $scGW1$ approach.  As described in the Supporting Information,
 in $scGW1$ the self consistent procedure is halted once a full unit of the self energy has been included, i.e.~the xc-potential is entirely replaced by self energy.  For this reason the QP energy shifts are quite similar to $G_0W_0$.  Furthermore, since the wavefunctions converge sooner than the energies within self consistent $GW$, the QP wavefunctions and vacuum level are also obtained within this procedure.  Indeed, the $scGW1$ spectra shown in Fig.~\ref{fgr:3} agree even better than $G_0W_0$ with the UPS and 2PP measurements  ($\Delta\varepsilon^{\textrm{UPS}} \approx -0.15$~eV, $\Delta\varepsilon^{\textrm{2PP}}\approx+0.04$~eV in Table \ref{tbl:1}).

\begin{figure}[!t]
\includegraphics[width=0.98\columnwidth]{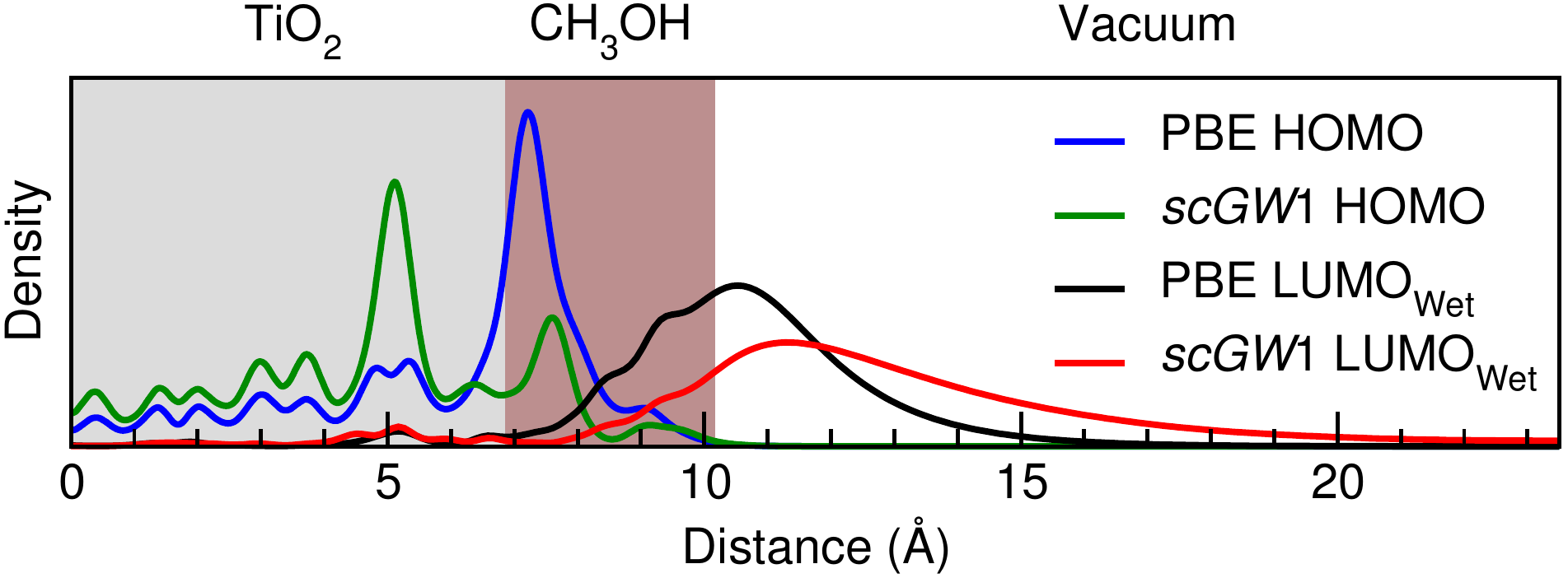}
\caption{
HOMO and LUMO$_{\textrm{Wet}}$ average density in the $x y$-plane versus distance in $\AA$ from the center of the TiO$_{\text{2}}$ substrate at $\Gamma$ as obtained from PBE (blue/black) and $scGW1$ (green/red). TiO$_{\text{2}}$ bulk, CH$_{\text{3}}$OH molecular layer, and vacuum are depicted by grey, brown, and white regions, respectively.}\label{fgr:4}
\end{figure}

To better understand the differences between the $G_0 W_0$ and $scGW1$ CH$_{\text{3}}$OH PDOS and Wet DOS shown in Figs.~\ref{fgr:1}(a) and \ref{fgr:3}, we consider the Kohn-Sham and QP HOMO and LUMO$_{\textrm{Wet}}$ wavefunctions depicted in Fig.~\ref{fgr:4}.  
We find the LUMO$_{\textrm{Wet}}$ level changes from being a molecular level with $\sigma^*$ character in PBE, to a more delocalized image potential-like level in $scGW1$, as was found previously for insulator surfaces \cite{ImageStatesLiFMgO}.  Just as the LUMO$_{\textrm{Wet}}$ level becomes more delocalized at the QP level, the HOMO becomes more hybridized with the three-fold coordinated oxygen atoms at the surface.  In fact, the screening is qualitatively different for molecular and bulk occupied levels.  This leads to a strong dependence of the QP energy corrections on the character of the occupied level.

\begin{figure}[!bt]
\includegraphics[width=\columnwidth]{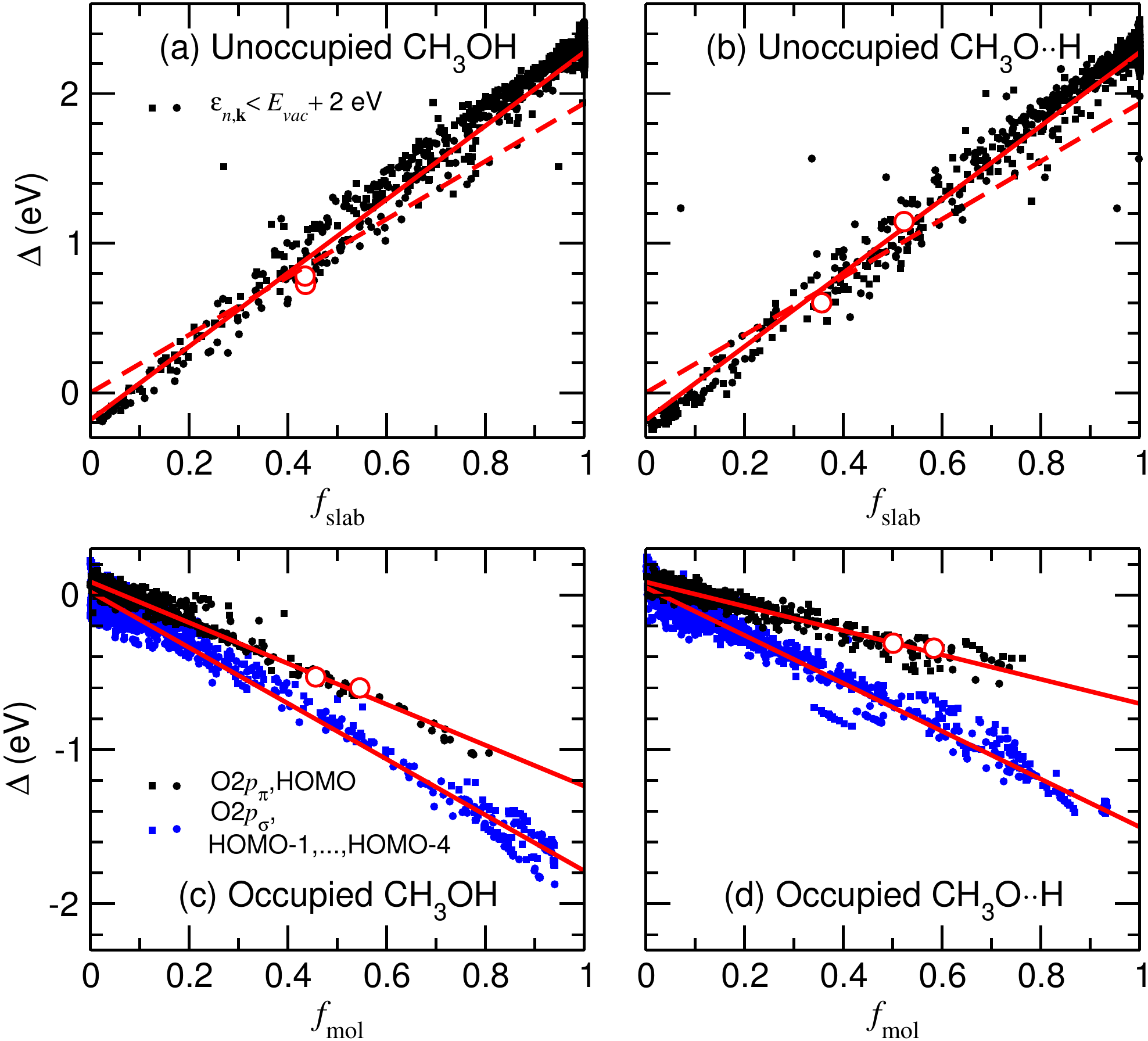}
\caption{
$G_0W_0$ QP energy correction $\Delta$ in eV versus fraction of the wavefunction's density (a,b) in the slab $f_{\textrm{slab}}$ for the unoccupied levels and (c,d) in the molecular layer $f_{\textrm{mol}}$ for the occupied levels 
of pairs I ($\bullet$) and II ($\blacksquare$) (a,c) intact and (b,d) half-dissociated structures. Open circles denote the (a,b) LUMO$_{\textrm{Wet}}$ and LUMO$_{\textrm{Wet}}+1$  and (c,d) HOMO at $\Gamma$ depicted in Fig.~\ref{fgr:1}.  Red solid lines are linear fits.  Dashed lines denote $\Delta\sim\Delta_{\textrm{bulk}}f_{\textrm{slab}}$.
}\label{fgr:2}
{\color{JACSBlue}{\rule{\columnwidth}{1.0pt}}}
\end{figure}

To determine the chemical origin of the QP energy corrections, we show in Fig.~\ref{fgr:2} how the shifts correlate with the bulk, molecular, and vacuum character of the wavefunction.  To quantify the wavefunction's character, we designate the bulk, molecular, or vacuum regions and calculate the fraction of the  wavefunction's surface averaged density in each, as shown in Fig.~\ref{fgr:4}.  
At the $G_0W_0$ level, the QP energy shift $\Delta$ is the difference between the QP self energy $\Sigma$ and the exchange-correlation potential $V_{xc}$, normalized by a factor $Z$, for a particular level, i.e.\ $\Delta \equiv Z(\Sigma - V_{xc})$ \cite{KresseG0W0,*KresseGW,*KressescGW,*CarusoPRB2012}.  We find that the QP shifts for the unoccupied levels correlate with the weight on the slab $f_{\textrm{slab}}$, which corresponds to the sum of the weights in the bulk (gray) 
and molecular (brown) regions. The same correlation for the unoccupied levels is obtained for all four structures.  Specifically, for the two intact and two half-dissociated structures we find the same linear correlation with $f_{\textrm{slab}}$,   as shown in Fig.~\ref{fgr:2}(a) and (b) respectively.  This indicates that the magnitude of the QP shift is determined by the fraction of the wavefunction that is not in the vacuum.  

Indeed, vacuum or free electron levels are reasonably well described by DFT.  This means the QP corrections are rather small, as the self energy of these levels is well described by their exchange-correlation potential, i.e.\ $\Sigma \sim V_{xc}$.  Because the wet electron levels have a large weight in the vacuum, their QP corrections are intermediate ($0.6$~eV $\lesssim\Delta\lesssim1.2$~eV) between bulk ($\Delta_{\textrm{slab}}\approx2.28$~eV) and vacuum levels ($\Delta_{\textit{vac}}\approx-0.18$~eV). From this we obtain the relation $\Delta \approx \Delta_{\textit{vac}}+(\Delta_{\textrm{slab}}-\Delta_{\textit{vac}})f_{\textrm{slab}}$ for the QP energy shift.

The QP corrections  for the LUMO$_{\textrm{Wet}}$ and LUMO$_{\textrm{Wet}}+1$ levels at $\Gamma$, which are depicted in Fig.~\ref{fgr:1}, are shown in Fig.~\ref{fgr:2}(a) and (b) as open circles.  For the half-dissociated geometry in Fig.~\ref{fgr:1}(b), we find the LUMO$_{\textrm{Wet}}$+1 level is significantly more hybridized with the bulk, compared to the LUMO$_{\textrm{Wet}}$ level, as well as the LUMO$_{\textrm{Wet}}$ and LUMO$_{\textrm{Wet}}$+1 levels of the intact structure in Fig.~\ref{fgr:1}(a).  As a consequence, the LUMO$_{\textrm{Wet}}$+1 level of the half-dissociated structure has a significantly larger QP shift, as seen in Fig.~\ref{fgr:2}(b).  

Fundamentally, we may justify the correlation between $\Delta$ and $f_{\textrm{slab}}$ by the following simple physical model.    First, because $\Sigma\sim V_{xc}$ for vacuum and nearly free electron levels, $\Delta_{\textit{vac}}\sim0$.  Second, because the QP energy shifts within the slab will be dominated by those of the bulk, we may approximate $\Delta_{\textrm{slab}}$ by the calculated QP shifts for bulk TiO$_{\text{2}}$ unoccupied levels, $\Delta_{\textrm{slab}}\sim\Delta_{\textrm{bulk}}\approx1.93$~eV.  From this we obtain the linear correlation $\Delta\sim\Delta_{\textrm{bulk}}f_{\textrm{slab}}$ with a standard deviation with respect to the full calculation of $\sigma\approx\pm0.2$~eV, as shown in Fig.~\ref{fgr:2}(a) and (b) as dashed lines. 
Altogether, these results suggest that $f_{\textrm{slab}}$ may serve as an effective descriptor for the QP energy shifts $\Delta$ of the unoccupied levels at the interface.

Because occupied levels are completely within the slab, $f_{\textrm{slab}}$ is an inappropriate descriptor for the shift of these levels.  For this reason we instead compare the QP shifts for the occupied levels of intact and half-dissociated methanol with the fraction of the wavefunction's density within the molecular layer $f_{\textrm{mol}}$.  Indeed, we find the QP corrections are well correlated with $f_{\textrm{mol}}$, as was the case for a NaCl insulating film on Ge(001) \cite{NaClCorrelation}.  This means  screening within the molecular layer plays an important role for occupied levels.  Moreover, the correlation is specific to the type of molecular layer, i.e.\ intact versus half-dissociated methanol.  

Further, we find the correlation is also orbital dependent, with separate correlations for the weakly-bonding O2$p_\pi$ and HOMO and the more strongly hybridized O2$p_\sigma$, HOMO$-1, \ldots ,$ and HOMO$-4$.  This is because it is easier to screen $\sigma$ orbitals, which are located between the atoms, than $\pi$ orbitals, which are out of plane.  Here the O2$p_\pi$ and O2$p_\sigma$ orbitals are labelled according to Ref.~\citenum{DuncanTiO2}\nocite{DuncanTiO2}, while HOMO$-1, \ldots,$ HOMO$-4$ include the corresponding levels for each type of methanol. The first set of levels includes the VBM and all levels close to the VBM in energy. For this reason, it is the relevant set of levels for hole trapping at the geometry of the unexcited methanol layer prior to photon irradiation. For the O2$p_\pi$ orbitals and focusing on the HOMO levels depicted in Fig.~\ref{fgr:1}, we find a larger QP shift for the intact ($\sim-1.2$~eV) versus the half-dissociated ($\sim-0.7$~eV) molecular layers, as shown in Fig.~\ref{fgr:2}(c) and (d), respectively, while bulk levels are essentially unshifted. 
  
The observed differences in the QP shifts are due to changes in the screening, as the xc-potential for HOMOs of the intact and half-dissociated methanol overlayers are almost the same.  These differences in screening are related to the proton associated charge transfer of 0.4 electrons shown in Fig.~\ref{fgr:0} for pairs I and II.  When charge is transferred out of the monolayer, its ability to screen electrons is reduced.  Further, the O2$p_\pi$ and HOMO levels are more strongly affected by charge transfer than $\sigma$ levels, as they are more easily polarized. This charge transfer dependence is the origin of the destabilization of the HOMO for the half-dissociated versus intact structures, which may explain the measured differences in photocatalytic activity. Such a destabilization of the HOMO has been previously observed by UPS for the catechol molecule on TiO$_{\text{2}}$(110) \cite{Diebold}.  
The same set of correlations hold for each type of structure from pair I and II.  Altogether, this means $f_{\textrm{mol}}$ is an appropriate descriptor for the QP energy shifts $\Delta$ of the occupied levels at the interface.

Overall, we find the screening of the occupied levels is affected by charge transfer (intact $\gtrsim$ dissociated) and the spacial distribution of the wavefunctions ($\sigma \gtrsim \pi$).  For the unoccupied levels, we find the correlation is independent of charge transfer.  This is because these levels are highly delocalized, so that minor changes in the local screening due to charge transfer are ``washed out'' by the larger differences in screening between the vacuum and slab regions.

In this work, many-body QP techniques have proven indispensable in obtaining a fundamental insight into the underlying processes which control the level alignment of methanol on TiO$_{\text{2}}$(110). 
Using the wavefunction character as a descriptor, we construct semi-quantitative models for predicting QP energies based on more practical DFT calculations. These models allow the study of a variety of properties that can affect the photocatalytic activity, such as the dynamical evolution of the level alignment, which are too costly to be computed directly at the QP level. These models are a major advancement in the accurate prediction of interfacial level alignment, which is of fundamental importance in photocatalysis. 

\titleformat{\section}{\bfseries\sffamily\color{JACSBlue}}{\thesection.~}{0pt}{\large$\blacksquare$\normalsize~}
\section*{ASSOCIATED CONTENT}
\subsubsection*{\includegraphics[height=8pt]{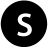} Supporting Information}
\noindent Energies, atomic coordinates, computational methods, PBE, $G_0 W_0$, $scGW1$, $scGW$, $scGW_0$, and HSE results.  This material is available free of charge via the Internet at http://pubs.acs.org.
\section*{AUTHOR INFORMATION}
\subsubsection*{Corresponding Authors}
\noindent *E-mail: annapaola.migani@cin2.es (A.M.).\\ *E-mail: duncan.mowbray@gmail.com (D.J.M.).\\ *E-mail: angel.rubio@ehu.es (A.R.).
\subsubsection*{Notes}
\noindent The authors declare no competing financial interest.

\section*{ACKNOWLEDGEMENTS}
We acknowledge funding from the European Projects DYNamo (ERC-2010-AdG-267374), CRONOS (280879-2 CRONOS CP-FP7); Spanish Grants (FIS2010-21282-C02-01, PIB2010US-00652, RYC-2011-09582, JAE DOC, JCI-2010-08156); Grupos Consolidados UPV/EHU del Gobierno Vasco (IT-319-07); NSFC (21003113 and 21121003); MOST (2011CB921404); and NSF Grant CHE-1213189; and computational time from i2basque, BSC Red Espanola de Supercomputacion, and EMSL at PNNL by the DOE.
\bibliography{Bibliography}

\end{document}